\documentclass[journal]{IEEEtran}
\usepackage{amssymb,amsmath,mathrsfs}
\usepackage{booktabs}
\usepackage{graphicx}  
\usepackage{subfigure}
\usepackage{epstopdf}
\usepackage{enumitem}
\usepackage{cite} 
\usepackage{nicefrac}
\usepackage{algorithm}
\usepackage{multirow}
\usepackage{setspace}
\usepackage{comment}
\usepackage{soul}
\usepackage[dvipsnames]{xcolor}

\usepackage{theorem}
\newtheorem{Assumption}{Assumption}
\newtheorem{Lemma}{Lemma}
\newtheorem{Theorem}{Theorem}
\newtheorem{Proposition}{Proposition}
\newtheorem{Remark}{Remark}

\allowdisplaybreaks[4]

\begin{document}

\title{\LARGE \bf Distributed Real-Time Cooperative Model Predictive Control} 

\author{Yuning Jiang, 
Kristína Fedorová, 
Roland Schwan,
Juraj Oravec, 
and Colin N. Jones

\thanks{The first two authors contributed equally. The authors acknowledge the Swiss National Science Foundation under the NCCR Automation grant 51NF40 180545, the Scientific Grant Agency of the Slovak Republic 1/0339/26, the Slovak Research and Development Agency APVV-24-0007, and NextGeneration EU Recovery and Resilience Plan Slovakia No.~09I01-03-V04-00024. YJ, RS, and CJ are with the Automatic Control Laboratory, EPFL. KF and JO are with the Slovak University of Technology in Bratislava.}

}

\maketitle

\begin{abstract}
This paper presents a novel distributed real-time iterative method for cooperative linear Model Predictive Control (MPC) with polyhedral state and input constraints, implementing a relaxed, recentered logarithmic barrier function. The method transforms centralized MPC problems into a distributed framework using temporal-spatial splitting, facilitating efficient online implementation. To achieve real-time performance, we introduce a distributed convex nonlinear optimization solver with guaranteed global linear convergence. By executing this solver with a fixed number of iterations per sampling time, the resulting near-optimal MPC controller ensures closed-loop stability while maintaining fixed computational complexity. The proposed approach is validated through an illustrative example involving a spring-vehicle-damper system benchmark, demonstrating both its effectiveness and computational efficiency.
\end{abstract}


\section{Introduction}
\label{sec:introduction}

As an advanced optimization-based technique, model predictive control (MPC) has been exhaustively investigated for its capability to deal with complex dynamic systems in both industry and academia~\cite{QIN2003733}. 
A recent trend in online solver design for large-scale systems, including the network-structured systems addressed in this paper, is to exploit a problem's structure by dividing it into smaller, more tractable subproblems~\cite{Donoghue2013,jiang2020parallel, Hult2022}. 
These approaches have been driven by significant research efforts in distributed optimization~\cite{Boyd2011}, such as the dual decomposition method~\cite{Everett1963}, the widely used Alternating Direction Method of Multipliers (ADMM)~\cite{Boyd2011}, and the Augmented Lagrangian-based Alternating Direction Inexact Newton (ALADIN) method~\cite{Houska2016}, having various successful applications in an MPC framework; see~\cite{Conte2012,Giselesson2013,Kozma2013,Necoara2008,Donoghue2013,Richter2011,jiang2020parallel}.
In cooperative MPC, these distributed online solvers are typically required to solve the problem optimally to guarantee closed-loop performance. However, even under mild conditions, achieving the desired accuracy to ensure performance is often impractical in large-scale scenarios due to the necessity of checking the stopping criterion at every iteration. This leads to significant communication overhead between local agents. The real-time constraints and limited sampling times further hinder the achievement of the desired accuracy.

To address these issues, considerable research has focused on simplifying the resulting optimization problems by incorporating \emph{a priori} constraint tightening~\cite{Rostami2023} or soft constraints~\cite{soft-constrained-MPC,soft-constrained-mpc-2,soft-constrained-mpc-3}.
These approaches benefit from converting complex constraint sets into simpler formulations, for instance by retaining only equality constraints or by relaxing hard constraints altogether.
However, such approximations may come at the cost of weakened or absent theoretical convergence guarantees. 
A perspective way to overcome this obstacle is introducing relaxed barrier functions, known as recentered relaxed logarithmic barrier (RRLB) functions, which have been extensively studied for stabilizing linear dynamical systems within an MPC design framework. Implementing RRLB functions allows the design of globally asymptotically stabilizing schemes, particularly through the proper choice of terminal costs and terminal sets~\cite{RRLB,RRLB-linear-MPC}. A recent decentralized real-time iteration scheme for distributed nonlinear MPC is proposed in \cite{stomberg2025}. However, for the considered MPC variant, the provided guarantees are only local.

This paper addresses the challenge of designing distributed real-time MPC for widely used large-scale, network-structured linear time-invariant systems with polyhedral constraints. The objective is to simplify the original complex real-time MPC problem into a manageable distributed form while ensuring closed-loop stability and recursive feasibility.  The proposed method introduces a novel temporal-spatial splitting scheme that achieves fixed complexity in both computational and communication overheads. Specifically, our approach requires only a fixed number of iterations per sampling time, eliminating the need to check stopping criteria at each iteration and thereby reducing communication overhead typical in existing distributed MPC schemes. This leads to a more efficient implementation, particularly in cooperative settings.

Section~\ref{sec:preliminaries} reviews the centralized RRLB-based MPC and the main idea of the ALADIN algorithm. The first contribution, established in Theorem~\ref{Thm::globallinear}, shows that ALADIN achieves a global linear convergence rate for distributed convex nonlinear programs with decoupled nonlinear objectives. Section~\ref{sec::RTI} introduces the main contributions. We first extend the centralized RRLB-based MPC to a distributed variant for network-structured systems by designing a separable terminal cost. Then, a real-time suboptimal MPC scheme is developed using ALADIN as the online solver, exploiting both spatial and temporal distributed structures. A precomputable stopping criterion based on a fixed number of online iterations is provided to ensure real-time closed-loop stability, following recent work~\cite{fedorova2023generalized}. 
Section~\ref{sec::case} validates the proposed controller, and Section~\ref{sec::conclusions} concludes this paper.

\textit{Notations:} For a matrix $\Sigma\in\mathbb S_{+}^{n}$ with $\Sigma\succ0$, we define $\|x\|_\Sigma=\sqrt{x^\top\Sigma x}$, where $\mathbb S_{++}^n$ ($\mathbb S_{+}^n$) denotes the set of positive (semi-)definite matrices in $\mathbb R^{n\times n}$. Moreover, we define $\mathbb Z_{z_1}^{z_2}=\{z\in\mathbb Z\mid z_1\le z\le z_2\}$. A block-diagonal matrix $S$ with blocks $S_i$, $i\in\mathcal N$, is denoted by $S=\mathrm{diag}_{i\in\mathcal N}(S_i)$ or $S=\mathrm{diag}(S_1,\ldots,S_m)$ with $m=|\mathcal N|$. We stack vectors $x_i$, $i\in\mathcal N$, as $x=\mathrm{col}_{i\in\mathcal N}(x_i)$ or $x=\mathrm{col}(x_1,\ldots,x_m)$.

\section{Preliminaries}
\label{sec:preliminaries}

This preliminary section has two parts. The first part recaps the main idea of the relaxed, recentered barrier-based MPC design, including its formulation and nominal stability properties. The second part recaps the ALADIN algorithm summarized in~\cite{Houska2021} and analyzes its global linear convergence property when applied to distributed convex nonlinear programs.

\subsection{Relaxed Recentered Logarithmic Barrier Based MPC}
\label{sec::RRBMPC}
Consider a linear time-invariant (LTI) system given by 
\begin{equation}
\label{eq::dynamic}
x_{k+1} = A x_k + Bu_k
\end{equation}
with state $x_k\in\mathbb R^{n_\mathrm{x}}$ and control inputs $u_k\in\mathbb R^{n_\mathrm{u}}$ at time instant $k$. Here, we consider state and input constraint sets
$\mathcal X:= \{ x\in\mathbb R^{n_{\mathrm x}}: G^\mathrm{x} x\leq d^{\mathrm x} \}$ and $\mathcal U:= \{ u\in\mathbb R^{n_{\mathrm u}}: G^\mathrm{u} u\leq d^{\mathrm u} \} $with $G^\mathrm{x}\in\mathbb R^{m_{\mathrm x}\times n_{\mathrm x}}$, $G^\mathrm{u}\in\mathbb R^{m_{\mathrm u}\times n_{\mathrm u}}$, $d^\mathrm{x}\in\mathbb R^{m_{\mathrm x}}_{++}$ and $d^\mathrm{u}\in\mathbb R^{m_{\mathrm u}}_{++}$, which are convex and closed polyhedra containing the origin in their strict interiors. In the context of MPC design, one conventional option to ensure the recursive feasibility is to introduce the soft constraints using the so-called relaxed recentered logarithmic barrier (RRLB) function.  This approach avoids the design-stage conservatism associated with \textit{a priori} constraint tightening, while potentially allowing limited transient constraint violations during iterations. To recap its main idea, we first define the barrier function $b_\delta:\mathbb R\to\mathbb R$ as % given by
\[
b_\delta(z)\; = \; \left\{
\begin{aligned}
-\ln(z),\qquad \qquad \qquad \qquad \quad &\quad \text{if } z>\delta, \\
\frac{1}{2}\left(
\left(
\frac{z-2\delta}{\delta}
\right)^2 - 1
\right)-\ln(\delta),&\quad  \text{if } z\leq \delta
\end{aligned}
\right.
\]
with $\delta >0$ leading to a convex, globally twice continuously differentiable function. Based on function $b_\delta(\cdot)$, we define the soft penalties for $\mathcal X$ and $\mathcal U$ by $\widetilde b(x,u)=\widetilde b_\mathrm{x}(x)+\widetilde b_\mathrm{u}(u)$ with
\begin{subequations}
\label{eq::barrier_cst} 
\begin{align}
\widetilde b_\mathrm{x}(x) =&\; \sum_{r=1}^{m_{\mathrm{x}}}w_{r}^\mathrm{x}\left(
b_\delta([-G^{\mathrm{x}}x+d^\mathrm{x}]_r)+\ln([d^\mathrm{x}]_r)
\right),\\
\widetilde b_\mathrm{u}(u)=&\;\sum_{r=1}^{m_{\mathrm{u}}}w_{r}^\mathrm{u}\left(
b_\delta([-G^{\mathrm{u}}u+d^\mathrm{u}]_r)+\ln([d^\mathrm{u}]_r)
\right).
\end{align}
\end{subequations}
and the weights $w^x \in \mathbb{R}^{m_\text{x}}_+$ and $w^u \in \mathbb{R}^{m_\text{u}}_+$. Hereafter, the dependence on $\delta$ is omitted for simplicity. Now, we introduce the RRLB-based MPC problem originally proposed in~\cite{Feller2018}
\begin{subequations}
\label{eq::rrbMPC}
\begin{align}
V(x_0):=\min_{x,u}\;\;& L^{\mathrm{f}}(x_{N})+\sum_{k=0}^{N-1} L(x_{k},u_{k})\\
\text{s.t.}\;\;\;& x_{k+1} = Ax_k + Bu_k, \;\forall\, k \in \mathbb{Z}_0^{N-1}.
\end{align}
\end{subequations}
Problem~\eqref{eq::rrbMPC} is an equality-constrained convex nonlinear problem that is parametric in $x_0$. Here, $x\in\mathbb R^{(N+1)n_\mathrm{x}}$ and $u\in\mathbb R^{N\cdot n_\mathrm{u}}$ stack state $x_k\in\mathbb R^{n_\mathrm{x}}$ for all $k\in\mathbb Z_0^{N}$ and control inputs $u_k\in\mathbb R^{n_\mathrm{u}}$ for all $k\in\mathbb Z_0^{N-1}$, respectively. The stage and terminal costs are given by 
\begin{equation}
\label{eq::costs} 
L(x,u) = l(x,u) + \xi\cdot\widetilde b(x,u)\;\text{ and }\;L^{\mathrm{f}}(x)=x^\top P x 
\end{equation}
with $l(x,u):= x^\top Q x + u^\top R u$, $Q, P\in \mathbb S_{++}^{n_{\mathrm x}}$, $R\in \mathbb S_{++}^{n_{\mathrm u}}$ and $\xi>0$. As discussed in~\cite[Section III.A]{Feller2018}, by properly choosing $w_r$ in~\eqref{eq::barrier_cst}, e.g., following the approach in~\cite{RRLB}, we can have $\nabla L(0,0) = 0$. According to Lemma~3 in~\cite{RRLB-linear-MPC}, the second-order derivative of $\widetilde b(\cdot,\cdot)$ is upper-bounded by
\begin{subequations}
\label{eq::logbar_bound}
\begin{align}
0\preceq\nabla_{x}^2\widetilde b(x,u)\preceq& \Delta Q=\frac{1}{2\delta^2}(G^\mathrm{x})^\top \text{diag}(\mathbf w^\mathrm{x})G^\mathrm{x},\\[0.12cm]
0\preceq\nabla_{u}^2\widetilde b(x,u)\preceq& \Delta R=\frac{1}{2\delta^2}(G^\mathrm{u})^\top \text{diag}(\mathbf w^\mathrm{u})G^\mathrm{u}
\end{align}    
\end{subequations}
such that we have 
\begin{equation}
\label{eq::extendedstage}
l(x,u)\leq L(x,u) \leq \; \widetilde l(x,u)=x^\top \widetilde Q x + u^\top \widetilde R u 
\end{equation}
with $\widetilde Q=Q+\xi\cdot\Delta Q$ and $\widetilde R=R+\xi\cdot\Delta R$.
\begin{Proposition}
\label{rem::nominal}
Let $(A, B)$ be controllable and the terminal cost $L^{\mathrm{f}}(\cdot)$ be a Lyapunov function, i.e., $L^{\mathrm{f}}(A_Kx) - L^{\mathrm{f}}(x) \leq L(x,Kx)$ with $K$ a pre-stabilizing feedback gain and $A_K=A+BK$, the RRLB-based MPC controller, obtained by recursively solving~\eqref{eq::rrbMPC}, asymptotically stabilizes system~\eqref{eq::dynamic}.
\end{Proposition}

This nominal stability property of the RRLB-based MPC has been exhaustively investigated in~\cite{RRLB-linear-MPC}. The key idea builds on the result that if the terminal cost is a Lyapunov function, the considered value function also exhibits properties of a Lyapunov function, thereby establishing closed-loop system stability~\cite{Rawlings2009}. A detailed proof refers to~\cite[Section IV.C]{RRLB-linear-MPC}.

\subsection{Distributed Convex Nonlinear Programming}
\label{sec::classicalALADIN}

Designing a fast and efficient online solver is the key to deploying the MPC controller in real-time scenarios. To this end, we propose using the ALADIN framework to solve online MPC problems, as it exhibits superior convergence properties compared to other relevant approaches, providing a linear convergence rate in the worst case, see~\cite{Houska2021}. Let us consider the following structured convex nonlinear programs
\begin{align}
\label{eq::disopt}
\min_{y} \! \; f(y) := \sum_{n=1}^{M}f_n(y_n) \; \; \text{s.t.} \; e = Dy := \sum_{n=1}^{M} D_ny_n \; \mid\lambda \, , 
\end{align}
with $y=[y_1^\top,...,y_{M}^\top]^\top$, $D=[D_1,...,D_{M}]$ for $M$ subsystems. Here, we use the notations $y_n$ to denote the $n$-th local vector of decision variables and $f_n(\cdot)$ its corresponding decoupled objective function. The couplings between $y_n$ are introduced by the affine equality constraints denoted by matrices $D_n$ and vector $e$. In the context of MPC design, the affine equality constraints are always given by linear dynamic equations~\eqref{eq::dynamic}. Hereafter, the corresponding Lagrangian multipliers are denoted right after the constraints, i.e., $\lambda$ denotes the Lagrangian multipliers of the coupled affine equality constraints. 

Algorithm~\ref{alg::ALADIN} outlines the main idea of the ALADIN method to solve~\eqref{eq::disopt}, including four main steps. Step~1) deals with augmented Lagrangian-based decoupled nonlinear problems (NLP)s~\eqref{eq::deNLP:general} in parallel. Step~2) defines the stopping criterion with $\bar{y} = \text{col}(\bar{y}_1,\ldots,\bar{y}_n)$. One can easily check if the conditions defined in Step~2) hold, the iteration $(y^\ell,\lambda^\ell)$ satisfies the first-order KKT optimality conditions up to a term of order $\mathcal O(\epsilon)$~\cite{Houska2021}. Based on the solution $y_n$ of decoupled Problems~\eqref{eq::deNLP:general}, Step~3) constructs a coordinated QP~\eqref{eq::coQP} with only equality constraints. 
In practice, one may employ distributed linear algebra routines to address~\eqref{eq::coQP} in a bilevel decentralized manner using only neighbor-to-neighbor communication~\cite{Engelmann2020}. However, such methods are inherently iterative and can only deliver inexact solutions with finite numerical accuracy, such that the proposed convergence analysis is no longer rigorously satisfied. Finally, Step 4) updates the primal-dual iterates based on the solution of Problem~\eqref{eq::coQP}. 
Note that the NLPs in~\eqref{eq::deNLP:general} have only nonlinear costs, hence differ from conventional nonlinear MPC design problems, e.g., see~\cite{GZ16}.
\begin{algorithm}[htbp!]
\caption{ALADIN method for solving Problem~\eqref{eq::disopt}.}
\textbf{Initialization:} select $\bar{y}^1,\lambda^1$, scaling matrices $H_n\succ 0$, $n\in\mathbb Z_1^{M}$, and $\epsilon>0$.\\
\textbf{Repeat:} 
\begin{enumerate}[leftmargin=12pt]
\item Solve decoupled NLPs in parallel for all $n\in\mathbb Z_1^{M}$,
\begin{equation}
\label{eq::deNLP:general}
\hspace{-7mm}y_n^\ell:=\text{arg}\min_{y_n}\;\;f_n(y_n) + (\lambda^\ell)^\top D_n y_n + \frac{1}{2}\left\|y_n-\bar y^\ell_n\right\|_{H_n}^2.
\end{equation}
\item Terminate if $\|Dy^\ell-e\|\leq\epsilon$ and $\|y^\ell-\bar y^\ell\|\leq\epsilon$ hold.
\item Solve coupled QP with communication
\begin{align}
\label{eq::coQP}      
\hspace{-7mm}\min_{\bar y}\sum_{n=1}^{M}\frac{1}{2}\left\|\bar y_n-2y_n^\ell+\bar y_n^\ell\right\|_{H_n}^2\; 
\text{s.t.}\,\sum_{n=1}^{M}D_n\bar y_n = e\;\mid\Delta \lambda.
\end{align}
\item Update $\bar y^{\ell+1} := \bar y$ and $\lambda^{\ell+1} := \lambda^\ell+\Delta \lambda$.
\end{enumerate}
\label{alg::ALADIN}
\end{algorithm}

In contrast to the conventional nonconvex ALADIN method introduced in~\cite{Houska2016}, the proposed Algorithm~\ref{alg::ALADIN} is tailored for the linear MPC design problem, i.e., handles a fixed scaling matrices $H_n$ in the least-square penalty term in the decoupled problem~\eqref{eq::deNLP:general} and the quadratic cost in the coupled QP~\eqref{eq::coQP}, cf.~\cite[Algorithm~2]{Houska2016}. In the following, we analyze the convergence property of Algorithm~\ref{alg::ALADIN} and establish its global linear convergence under the following assumptions:
\vspace{-0.1cm}
\begin{Assumption}
\label{ass::aladin}
We assume that
\begin{enumerate}[leftmargin=15pt]
\item[a)] Functions $f_n$ in~\eqref{eq::deNLP:general} are closed, proper, and twice continuously differentiable. Moreover, they are strongly convex and have a Lipschitz continuous gradient, i.e., there exist matrices $\underline H_n, \overline H_n \succ 0$ such that $\underline H_n \preceq \nabla^2f_n \preceq \overline H_n $ holds;
\item[b)] $D=[D_1,...,D_{M}]$ in~\eqref{eq::deNLP:general} is full row rank, i.e., the Linear Independent Constraint Qualification (LICQ) is satisfied. 
\end{enumerate}
\end{Assumption}

Then, we introduce a Lyapunov-type function to evaluate the distance between the current iterate and the optimal solution as follows, $\Omega\left(y,\lambda\right) = \left\|y-y^\star\right\|_H^2 + \left\|\lambda-\lambda^\star\right\|^2_\Sigma$ with $H=\text{diag}(H_n)$, $\Sigma = D H^{-1}D^\top $, and $(y^\star,\lambda^\star)$ the optimal solution of~\eqref{eq::disopt}, to establish the global linear convergence.
\begin{Theorem}
\label{Thm::globallinear}
Let Assumption~\ref{ass::aladin} hold and let Problem~\eqref{eq::disopt} be feasible. The primal-dual iterations $(\bar y^\ell,\lambda^\ell)$ of Algorithm~\ref{alg::ALADIN} converge to the (global) optimum solution $(y^\star,\lambda^\star)$ with a linear convergence rate. Simultaneously, there exists a constant $\gamma$ such that the linear convergence rate $\kappa = \frac{\gamma^2}{4+\gamma^2}<1$ satisfies
\begin{equation}
\label{eq::linearcontraction}
\Omega(\bar y^{\ell+1} ,\lambda^{\ell+1})\leq \kappa\cdot \Omega(\bar y^\ell, \lambda^\ell) \, .
\end{equation}
\end{Theorem}

The proof of Theorem~\ref{Thm::globallinear} is derived in Appendix~\ref{app::convergence}, following two main steps. First, based on Theorem~1 in~\cite{houska2017convex}, we have 
\[
\Omega(\bar y^{\ell+1} , \lambda^{\ell+1})+4\Phi(y^\ell-y^\star)\leq \Omega(\bar y^\ell, \lambda^\ell)
\]
with $\Phi(\cdot) = \|\cdot\|_{\underline{H}}^2$.
Then, we can prove that there exists a constant $\gamma$ such that  
$\Omega(\bar y^{\ell+1} , \lambda^{\ell+1})\leq \gamma^2\cdot\Phi(y^{\ell}-y^\star)$.
Combining these two inequalities results in the linear contraction below, which can conclude the proof as follows, $\Omega(\bar y^{\ell+1} , \lambda^{\ell+1})\leq \kappa\cdot\Omega(\bar y^\ell, \lambda^\ell)\;\;\text{with}\;\;\kappa=\frac{\gamma^2}{4+\gamma^2} < 1$\,.

\section{Real-Time Distributed MPC Scheme}
\label{sec::RTI}  
This section presents the main contribution of the paper: a real-time distributed MPC scheme for network-structured linear time-invariant systems. We first extend the centralized RRLB-based MPC formulation to network-structured settings and design a separable terminal cost that preserves nominal stability in the distributed scheme. Finally, an ALADIN-based distributed real-time control approach is proposed.

\subsection{Network Structured LTI Systems} 
Let us consider the LTI system~\eqref{eq::dynamic} with block-wise system coefficient matrices given by $A=[A_{i,j}]_{i,j=1}^M$, $A_{i,j} =0$ if $(i,j)\notin \mathcal E$, and $B=\text{diag}(B_1,...,B_M)$, states $x_k = [x_{1,k}^\top,...,x_{M,k}^\top]^\top$ and control inputs $u_k = [u_{1,k}^\top,...,u_{M,k}^\top]^\top$ that stack the local system states and control inputs of $M$ subsystems, respectively. Here, we use the tuple $(\mathcal N, \mathcal E)$ to denote the topology of the system, where $\mathcal N$ is the index set of the subsystem with $|\mathcal N| = M$ and $\mathcal E$ denotes the pairs of neighbors, i.e., $\mathcal E\subseteq \mathcal N\times\mathcal N$. For each subsystem $i$,
\begin{equation}
\label{eq::localdyanmic}
x_{i,k+1}  = \sum_{j\in\mathcal N_i}A_{i,j}x_{j,k} + B_iu_{i,k}= A_{\mathcal N_i}x_{\mathcal N_i,k} + B_i u_{i,k}
\end{equation}
holds with $x_{i,k}\in\mathbb R^{n_{\mathrm{x}_i}}$ and $u_{i,k}\in\mathbb R^{n_{\mathrm{u}_i}}$ the system state and control input of the $i$-th subsystem at time instant $k$. Set $\mathcal N_{i}$ collects the indexes of the neighbors of subsystem $i$ including itself, i.e., if $(i,j)\in\mathcal E$ then, $j\in\mathcal N_i$ while $i\in\mathcal N_i$. The matrices $A_{i,j}\in\mathbb R^{n_{\mathrm{x}_i}\times n_{\mathrm{x}_j}}$ are the coupling system matrices between the subsystem $i$ and $j$. Without loss of generality, we focus on the control setup where the neighbors' coupling is only present in the state space domain but not in the control inputs.  
Moreover, we consider the state and input constraint sets $\mathcal X: = \mathcal X_1\times \cdots \times \mathcal X_M $ and $\mathcal U:=\mathcal U_1\times \cdot \times \mathcal U_M$ specific to each subsystem, and let the following assumption holds.
\begin{Assumption}
\label{ass::sets}
We assume that $(A, B)$ in~\eqref{eq::dynamic} is controllable, $0 \in relint(\mathcal{U}_i)$ and $0 \in relint(\mathcal{X}_i)$, where relint($\cdot$) denotes the relative interior of a set.
\end{Assumption}  

\subsection{Separable Terminal Cost Design}
\label{sec::sec::terminalcost}
Based on Proposition~\ref{rem::nominal}, one can see the key idea to enforce closed-loop stability of the nominal RRLB-based MPC controller is to design a terminal cost which is a Lyapunov function. First, we introduce the following assumption from~\cite{Conte2016}.
\begin{Assumption}
\label{ass::separable_control_law}
There exists a linear state-feedback control law $\mu(x) = Kx = \mathrm{col}_{i\in \mathcal N}(K_{\mathcal{N}_i}x_{\mathcal{N}_i})$
such that the LTI system in~\eqref{eq::dynamic} is asymptotically stable for $u_k = \mu(x_k)$.
\end{Assumption}
Here, $x_{\mathcal N_i}$ stacks $x_j$ for all $j\in\mathcal N_i$ while the $K$ matrix is block-wise with the network structure embedded, i.e., block $K_{i,j}\neq 0$ for all $j\in\mathcal N_i$ and $K_{i,j}=0$ otherwise. Thus, $K_{\mathcal N_i}$ collects the non-zero blocks $K_{i,j}$. 
Next, to simplify notation, the corresponding selection matrices $T_i\in\{0,1\}^{n_{\mathrm{x}_i}\times n_\mathrm{x}}$ and $W_i\in\{0,1\}^{n_{\mathrm{x}_{\mathcal N_i}}\times n_\mathrm{x}}$ with $x_i=T_ix$ and $x_{\mathcal N_i} = W_ix$. Based on Assumption~\ref{ass::separable_control_law}, we recall the following result. 
\begin{Proposition}[Theorem~6 in~\cite{Conte2016}]
\label{prop::Lyapunov_terminal}
Given a structured control law $\mu(\cdot)$ satisfying Assumption~\ref{ass::separable_control_law}. If there exist functions $\vartheta_i(\cdot)$ such that the following conditions hold,
\begin{subequations}
\label{eq::Lyapunov_term}
\begin{align}\label{eq::localLyapunov}
L^{\mathrm{f}}_i(x_i^+)-L^{\mathrm{f}}_i(x_i)&\leq -\widetilde l_i(x_{i}, K_{\mathcal{N}_i}(x_{\mathcal{N}_i}))+\vartheta_i(x_{\mathcal{N}_i}),\\\label{eq::residue}
\sum_{i=1}^M\vartheta_i(x_{\mathcal{N}_i})&\leq0,
\end{align}
\end{subequations}
with $x_{i}^+ = A_{\mathcal N_i}x_{\mathcal N_i}+Bu_{k}$, then 
$L^{\mathrm{f}}(x) = \sum_{i=1}^ML^{\mathrm{f}}_i(x_i)$ is a Lyapunov function for LTI system in~\eqref{eq::dynamic} for $u_{k} = \mu(x_{k})$.
\end{Proposition}
\vspace{-0.1cm}

Based on Proposition~\ref{prop::Lyapunov_terminal}, we tailor the design scheme proposed in~\cite{Conte2016} to construct the separable terminal cost $L^{\mathrm{f}}_i(\cdot)$ satisfying the conditions~\eqref{eq::Lyapunov_term}. First, we define non-negative function $\vartheta_i(x_{\mathcal N_i})=x_{\mathcal N_i}^\top \Gamma_{\mathcal{N}_i}x_{\mathcal N_i}$ and next, we design matrices $P_i$ and $\Gamma_{\mathcal N_i}$, and feedback gains $K_{\mathcal N_i}$ by solving the following convex semidefinite program (SDP) 
\begin{equation}
\label{eq::LMI}
\min_{E_i,F_{\mathcal N_i}, E_{\mathcal N_i},Y_{\mathcal N_i}}\;\;-\sum_{i=1}^M\log \det(E_i)
\end{equation}
% subject to $\forall\,i\in\mathcal N$, 
subject to \quad $\sum_{i=1}^M W_i^\top F_{\mathcal N_i} W_i \preceq 0$, \quad $\forall\,i\in\mathcal N$,
\begin{align*}
&
{\footnotesize
\setlength\arraycolsep{1.5pt}
\begin{bmatrix}
\overline E_i+F_{\mathcal N_i}& E_{\mathcal N_i}A_{\mathcal N_i}^\top + Y_{\mathcal N_i}^\top B_i^\top &E_{\mathcal N_i}\mathcal Q_i^{1/2} & Y_{\mathcal N_i}^\top R_i^{1/2}\\[0.12cm]
A_{\mathcal N_i}E_{\mathcal N_i} +B_iY_{\mathcal N_i}& E_i & 0 & 0\\[0.12cm]
\mathcal Q_i^{1/2}E_{\mathcal N_i} & 0 & I & 0 \\[0.12cm]
R_i^{1/2}Y_{\mathcal N_i} & 0 & 0 & I 
\end{bmatrix} \succeq  0 \, , } 
\end{align*}
with $\mathcal Q_i=\partial^2_{x_{\mathcal N_i}} \tilde{l}_i(x_i,u_i)$, $E_i = P_i^{-1}$, $\overline E_i = W_iT_i^\top E_i T_i W_i^\top$, $E_{\mathcal N_i} = W_iEW_i^\top $, 
$F_{\mathcal N_i}=E_{\mathcal N_i}\Gamma_{\mathcal N_i} E_{\mathcal N_i}$, and $Y_{\mathcal N_i}= K_{\mathcal N_i} E_{\mathcal N_i}$ for all $i\in\mathcal N$. Since~\eqref{eq::LMI} is derived from~\eqref{eq::Lyapunov_term}, the feasibility of~\eqref{eq::LMI} guarantees that~\eqref{eq::Lyapunov_term} holds. 
We point out that the numerical/runtime complexity associated with solving the SDP in~\eqref{eq::LMI} does not directly impact the real-time runtime of the MPC controller, as the procedure proposed to solve the problem in~\eqref{eq::LMI} is executed just once and offline.

Based on Remark~\ref{rem::nominal} and Proposition~\ref{prop::Lyapunov_terminal}, we can extend the nominal closed-loop system stability condition for RRLB-based MPC in~\cite{RRLB-linear-MPC} to network dynamics~\eqref{eq::localdyanmic} with separable terminal cost based on the design procedure introduced above. The result is summarized in the following theorem. 
\vspace{-0.2cm}
\begin{Theorem}
\label{thm::nominalstability}
Let Assumptions~\ref{ass::sets},~\ref{ass::separable_control_law} hold and let~\eqref{eq::LMI} be feasible. \textcolor{black}{Given the separable terminal cost designed by solving ~\eqref{eq::LMI} such that conditions~\eqref{eq::Lyapunov_term} hold.}
For any initial point $x_0\in\mathbb R^{n_\mathrm{x}}$, the MPC controller obtained by optimally solving~\eqref{eq::rrbMPC}, with separable state and terminal costs given by 
$L(x,u):=\sum_{i=1}^M L_i(x_i,u_i),\;
L^{\mathrm{f}}(x):=\sum_{i=1}^ML^{\mathrm{f}}_i(x_i)$ 
asymptotically stabilizes the LTI system~\eqref{eq::dynamic} to the origin.
\end{Theorem}
Here, the separable stage and terminal costs indicate that in cost function~\eqref{eq::costs}, we have $Q=\text{diag}\{Q_i\}$, $R=\text{diag}\{R_i\}$, and $P=\text{diag}\{P_i\}$ with $Q_i,P_i\in\mathbb S_{++}^{n_{\mathrm x_i}}$ and $R_i\in \mathbb S_{++}^{n_{\mathrm u_i}}$.
The proof of Theorem~\ref{thm::nominalstability} can be found in Appendix~\ref{app::nominal}. Notice that this nominal stability relies on solving Problem~\eqref{eq::rrbMPC} to optimum. However, in real-time practice, the hardware platform may have computational capability, or the sampling time may be limited due to the fast system dynamics.
\vspace{-0.3cm}
\subsection{Real-Time Distributed MPC Scheme}
\label{sec::algorithm}
In the remainder of this section, we will propose a novel real-time scheme to solve Problem~\eqref{eq::rrbMPC} for network system~\eqref{eq::localdyanmic} approximately while the closed-loop asymptotic stability condition is still preserved. First, we introduce compact shorthands for notational simplification. Hereafter, the local decision variables with respect to the $i$-th subsystem at time instant $k$ are summarized as
$y_{i,0}=u_{i,0}$, $y_{i,k}=[x_{i,k}^\top$, $u_{i,k}^\top]^\top$ for all $k \in \mathbb{Z}_{1}^{N-1}$, and $y_{i,N}=x_{i,N}$, the corresponding costs $f_{i,k}(y_{i,k})=L_i(x_{i,k},u_{i,k})$,  and $f_{i,N}(y_{i,N})=L^{\mathrm{f}}_i(x_{i,N})$. Then, we introduce the compact formulation for the local dynamic equations~\eqref{eq::localdyanmic} as
\begin{equation}
\label{eq::densedynamic}
\begin{array}{c}
\mathcal C_{i,k}y_{i,k+1} + \sum_{j\in\mathcal N_i}\mathcal A_{i,j,k}\, y_{j,k} = h_{i,k},
\end{array}
\end{equation}
where for all $i\in\mathcal N$, $h_{i,0}=-A_{\mathcal N_i}x_{\mathcal N_i,0}$, $\mathcal A_{i,i,0}=B_i$, $\mathcal A_{i,j,0}=0$ for all $j\in\mathcal N_i\backslash\{i\}$, and $\mathcal C_{i,k}=[-I\;0]$, $\;k\in\mathbb Z_0^{N-1}$, and
\[
\mathcal A_{i,i,k} = [A_{i,i}\;B_i],\;h_{i,k}=0,\;
\mathcal A_{i,j,k}=[A_{i,j}\;0],\;j\in\mathcal N_i\backslash\{i\}.
\]

Now, we rewrite Problem~\eqref{eq::rrbMPC} into a distributed form
\begin{subequations}
\label{eq::mpc_dense}    
\begin{align}
V(x_0)& = \min_{y}\;\;\;\sum_{i=1}^M\sum_{k=0}^Nf_{i,k}(y_{i,k})\\
\text{s.t.}\;&\left\{
\begin{aligned}
&\mathcal C_{i,k}y_{i,k+1} + \sum_{j\in\mathcal N_i}\mathcal A_{i,j,k} y_{j,k} = h_{i,k},\;\;\mid \lambda_{i,k},\\
& \forall\, i \in \mathcal N,\; \forall\, k \in \mathbb{Z}_0^{N-1} .
\end{aligned}
\right.
\end{align}
\end{subequations}
\begin{Remark}
As functions $f_{i,k}$ for all $k\in\mathbb Z_0^{N-1}$ is a sum of the strongly convex quadratic stage cost and the RRLB function, we have $f_{i,k}$ satisfy Assumption 1.a) following the fact that the RRLB function is convex and twice continuously differentiable with bounded Hessian~\eqref{eq::logbar_bound}. Moreover, we have the coupled affine equation constraints in~\eqref{eq::mpc_dense}, given by the linear dynamic equations such that Assumption 1.b) holds.
\end{Remark}

For~\eqref{eq::mpc_dense}, the corresponding Lagrangian function is given by 
\begin{align*}
\mathcal L(y,\lambda) = \;&\sum_{i=1}^M \left(\mathcal L_{i,N}(y_{i,N},\lambda_{i,N-1})+ \mathcal L_{i,0}(y_{i,0},\lambda_{\mathcal N_i,0})\right)   \\
&+\sum_{i=1}^M\sum_{k=1}^{N-1}\mathcal L_{i,k}(y_{i,k},\lambda_{\mathcal N_i,k},\lambda_{i,k-1}).     
\end{align*}
Here, $\lambda_{\mathcal N_i,k}$ stacks $\lambda_{j,k}$ for all $j\in\mathcal N_i$ similar to the notation of $x_{\mathcal N_i,k}$, and the separable Lagrangian $\mathcal L_{i,k}$ are given by
\begin{align*}
&\mathcal L_{i,0}(y_{i,0},\lambda_{\mathcal N_i,0}) =f_{i,0}(y_{i,0})  +y_{i,0}^\top \sum_{j\in\mathcal N_i} \mathcal A_{j,i,k}^\top \lambda_{j,0}+h_{i,0}^\top \lambda_{i,0}, \\
&\mathcal L_{i,N}(y_{i,N},\lambda_{i,{N-1}}) = f_{i,N}(y_{i,N}) + \lambda_{i,N-1}^\top \mathcal C_{i,N-1} y_{i,N},
\end{align*}
and $\mathcal L_{i,k}(y_{i,k},\lambda_{\mathcal N_i,k},\lambda_{i,k-1}) =
f_{i,k}(y_{i,k})+ \lambda_{i,k-1}^\top\mathcal C_{i,k-1}y_{i,k}+y_{i,k}^\top\sum_{j\in\mathcal N_i}  \mathcal A_{j,i,k}^\top \lambda_{j,k}$. Before introduce the main algorithm, we define the shorthand $\Psi(y) = \sum_{i=1}^M\sum_{k=0}^N \Psi_{i,k}(y_{i,k})$, $\Psi_{i,0}(y_{i,0})=y_{i,0}^\top R_iy_{i,0}$ and $\Psi_{i,k}(y_{i,k})=y_{i,k}^\top H_i y_{i,k}$ with $H_i = \text{diag}(Q_i, R_i)$ and $\Psi_{i,N}(y_{i,N})=L^{\mathrm{f}}_i(y_{i,N})$. 
Here, $H_i$ is chosen as a tractable approximation of the Hessian from~\eqref{eq::mpc_dense}.

Algorithm~\ref{alg::aladin2} outlines the proposed real-time distributed MPC scheme, which builds upon the framework established in Algorithm~\ref{alg::ALADIN}, with an online routine in Step 2). Instead of checking the stopping criterion in Step 2) of Algorithm~\ref{alg::ALADIN}, in Algorithm~\ref{alg::aladin2} we execute a fixed number of iterations in Step 2), ensuring a predictable computational load necessary for real-time implementation. In Step~2a), corresponding to Step~1 of Algorithm~1, $M \cdot (N+1)$ unconstrained NLPs are solved in parallel, significantly reducing computational demand. Compared to earlier approaches with a primal focus just on the decomposition in the time domain, this formulation also introduces distribution in the state-space domain, further reducing the size of individual subproblems. Step~2b), which reflects the role of Step~3) in Algorithm~1, then solves a coupled LQR-like Problem~\eqref{eq::cLQR} to track the reference trajectory $(2y^\ell - \bar y^\ell)$, constructed from the local solution $y^\ell$ and the current primal iterate $\bar y^\ell$. Benefiting from its LQR-like structure, Problem~\eqref{eq::cLQR} is efficiently solved with minimum computational effort. The primal solution $\bar y^{\ell+1}$ from~\eqref{eq::cLQR} updates the primal iterate, while the dual iterate $\lambda^{\ell+1}$ is updated based on the dual solution increment $\Delta \lambda^{\ell}$ from~\eqref{eq::cLQR}. The efficient decomposition of the original problem into small-scale NLPs and the evaluation of parallel computing nominate the proposed control method for real-time implementation, even on industrial/embedded hardware with limited resources.
\vspace{-0.4cm}
\begin{algorithm}[hbpt!]
\caption{Distributed real-time control step evaluation for cooperative MPC.}
%\small
\textbf{Initialization:} select a initial guess $(\bar y^1, \lambda^1)$, a constant $\alpha > 0$, and a maximum number $\bar{\ell}$ of iterations.\\
\textbf{Online:}
\begin{enumerate}[leftmargin=12pt]
\item Obtain state measurement $x(t)$ and evaluate
\(\Gamma^1 = \|\bar{y}^1\|_H^2+\|\lambda^1\|_{\Sigma}^2.\)
If $\Gamma^1 > \alpha^2\|x(t)\|_Q^2$, rescale initial guesses 
\[
\bar y^1 := \bar y^1\sqrt{\alpha^2\|x(t)\|_Q^2/\Gamma^1}, \quad \lambda^1 := \lambda^1\sqrt{\alpha^2\|x(t)\|_Q^2/\Gamma^1}.
\]
\item[2)] \textbf{For:} $\ell \rightarrow \bar{\ell}$
\begin{enumerate}
\item Solve decoupled convex NLPs in parallel 
\begin{subequations}
\label{eq::deNLP}
\label{eq::ALADIN:STEP2}
\begin{align}\label{eq::1stdeNLP}
\hspace{-5mm}\min_{y_{i,0}}&\;\mathcal L_{i,0}(y_{i,0},\lambda_{\mathcal N_i,0}^\ell)+\Psi_{i,0}(y_{i,0}-\bar y_{i,0}^\ell ),\\\label{eq::kthdeNLP}
\hspace{-5mm}\min_{y_{i,k}}&\;\mathcal L_{i,k}(y_{i,k},\lambda_{\mathcal N_i,k}^\ell,\lambda_{i,k-1}^\ell) +\Psi_{i,k}(y_{i,k}-\bar y_{i,k}^\ell),\\\label{eq::NthdeNLP}
\hspace{-5mm}\min_{y_{i,N}}&\;\mathcal L_{i,N}(y_{i,N},\lambda_{i,N-1}^\ell)+\Psi_{i,N}(y_{i,N}-\bar y_{i,N}^\ell) ,
\end{align}
\end{subequations}
\item[] to achieve solutions $y_{i,k}^\ell$, $i\in\mathcal N$, $k\in\mathbb Z_1^{N-1}$.
\item Solve coupled QP problem
\begin{align}\label{eq::cLQR}
\hspace{-8mm}\min_{\bar y}\;& \sum_{i=1}^M \sum_{k=0}^N \frac{1}{2}\Psi_{i,k}(\bar y_{i,k}-2y_{i,k}^\ell+\bar y_{i,k}^\ell)\\\notag
\hspace{-8mm}\text{s.t.}\;&\mathcal C_{i,k}\bar y_{i,k+1} + \sum_{j\in\mathcal N_i}\mathcal A_{i,j,k} \bar y_{j,k} = h_{i,k}\mid \Delta \lambda_{i,k},\,\forall i,k%k\in\mathbb Z_0^{N-1},\,i\in\mathcal N
\end{align}
\vspace{-0.3cm}
\item[] update $\bar{y}^{\ell+1}:=\bar{y}$ and $ 
\lambda^{\ell+1}:=\lambda^{\ell}+\Delta\lambda^{\ell}$.
\end{enumerate}
\item[3)] Apply $u_0^{\bar{\ell}} = \text{col}_{i\in\mathcal{N}}(y_{i,0}^{\bar{\ell}})$ to the system in~\eqref{eq::dynamic} .
\item[4)] For all $i\in\mathcal{N}$, update $\bar y^1_i := [y_{i,1}^{\bar{\ell}\top},\ldots,y_{i,N}^{\bar{\ell}\top},y_{i,N}^{\bar{\ell}\top}]^\top$ and $\lambda^1_i := [\lambda_{i,1}^{\bar{\ell}\top},\ldots,\lambda_{i,N-1}^{\bar{\ell}\top},\lambda_{i,N-1}^{\bar{\ell}\top}]^\top$, go to Step~1.
\end{enumerate}
\label{alg::aladin2}
\end{algorithm}
\vspace{-0.2cm}

The rationale behind the rescaling in Step~1) of Algorithm~\ref{alg::aladin2} is that this step prevents initialization that is beyond a considerable distance from 0. If the term $\|\bar{y}^1\|_H^2 + \|\lambda^{1}\|^2_\Sigma$ is much larger than $\|x(t)\|_Q^2$, then $(\bar{y}^1, \lambda^1)$ are expected to be far away from the optimal solution $(y^\star, \lambda^\star)$ and it is worth rescaling these variables such that they have a reasonable order of magnitude to improve the convergence properties, i.e., to reduce the number of iterations $\ell$.

\subsubsection{Closed-Loop Stability Analysis}
First, we assume that:
\vspace{-0.2cm}
\begin{Assumption}
\label{ass::alpha}
There exists constant $\alpha > 0$ such that $\forall t>0$, $\|y^\star(x(t))\|^2_H+\|\lambda^\star(x(t))\|^2_\Sigma \leq \alpha^2 \|x(t)\|_Q^2$ with matrices $H$ and $\Sigma$ defined in function $\Omega(\cdot,\cdot)$.
\end{Assumption}
\vspace{-0.2cm}

Then, before we construct the closed-loop stability of the controller introduced in Alg.~\ref{alg::aladin2}, we need the following result. 
\vspace{-0.2cm}
\begin{Lemma}
\label{Lem::value}
Based on the proposed MPC iteration outlined in Algorithm~\ref{alg::aladin2}, for any $x_0\in\mathbb R^{n_\mathrm{x}}$,
there exist constants $\eta_1,\eta_2>0$ satisfying 
$|V(x_1^\star)-V(x_0^+)|\leq\eta_1 \|x_0\|_Q\|x_1^\star-x_0^+\|_Q+\eta_2 \|x_1^\star-x_0^+\|_Q^2$
with $x_1^\star=Ax_0+Bu_0^\star$ and $x_0^+=Ax_0+Bu_0^{\bar\ell}$. 
\end{Lemma}
\vspace{-0.4cm}
The proof is given in Appendix~C. The following theorem establishes the condition under which the proposed controller achieves asymptotic stability.
\vspace{-0.2cm}
\begin{Theorem}
\label{thm::m_bound}

Let Assumptions~\ref{ass::sets}--\ref{ass::alpha} hold and there exist $\sigma>0$ such that inequality $B_i^\top Q_iB_i\preceq \sigma R_i$ holds for all $i \in \mathcal{N}$. 
Then, the closed-loop system governed by the control law $u_0^{\bar{\ell}} = \mathcal{S} y^{\bar{\ell}}$, with $\mathcal{S} = [B,0,\ldots,0]$, given by Algorithm~2, is asymptotically stable for any $\bar{\ell} \in \mathbb{N}$ with
\begin{equation}
\label{eq::mbarBound}
\bar{\ell} > \frac{2\log\left(2\eta_1\alpha\sqrt{\sigma(1+\kappa)/\kappa}+4\eta_2\alpha^2\sigma(1+\kappa)/\kappa\right)}{\log(1/\kappa)}.
\end{equation}
\end{Theorem}

The proof refers to Appendix~\ref{app::m_bound}, where the analysis follows the main idea in~\cite{fedorova2023generalized}. However,~\cite{fedorova2023generalized} only considered the case of global linear convergence of the primal iteration, while we extend it in this paper to the primal-dual iteration. Note, constants $\sigma$, $\eta_1$, $\eta_2$, and $\kappa < 1$ in Theorem~\ref{thm::m_bound} depend on the problem data only, i.e., are independent of the initial state $x_0$.
\vspace{-0.6cm}
\begin{Remark}
The lower bound~\eqref{eq::mbarBound} monotically increases with $\sigma$, $\eta_1$, $\eta_2$, and $\kappa$. Thus, lower values of these constants yield a tighter bound. However, obtaining $\eta_1$, $\eta_2$ with reasonable effort is challenging, even offline, necessitating the use of conservative estimates. Consequently, such estimates result in a conservative lower bound~\eqref{eq::mbarBound}. Nevertheless, $\eta_1$, $\eta_2$ can be approximated using methods from approximate dynamic programming~\cite{Wang2015}. Note that our main point here is that we can find a finite bound on the number of iterations such that asymptotic stability can be guaranteed.
\end{Remark}

\subsubsection{Maximal Constraint Violation Guarantees} If~\eqref{eq::mpc_dense} is optimally solved online, we have inequalities 
$G^\mathrm{x}x_1^{\star} \leq d^\mathrm{x} + \hat z_\mathrm{x}(x_0,\delta)$ and $G^\mathrm{u}u_0^{\star} \leq d^\mathrm{u} + \hat z_\mathrm{u}(x_0,\delta),$
hold~\cite[Theorem~6]{RRLB-linear-MPC}
with the $i$-th element of $\hat z_\mathrm{x}(x_0,\delta)\in\mathbb R^{n_\mathrm{x}}$ given by $[\hat z_\mathrm{x}(x_0,\delta)]_i:=$
\[
\max_{x}\left\{[G^\mathrm{x}]_ix-[d^\mathrm{x}]_i\left|\widetilde b_\mathrm{x}(x)\leq \left(V(x_0)-x_0^\top P_\mathrm{uc}^* x_0\right)/{\xi}\right\}\right.,
\]
and $[\hat z_\mathrm{u}(x_0,\delta)]_i:=$
\[
\max_{u}\left\{[G^\mathrm{u}]_iu-[d^\mathrm{u}]_i\left|\widetilde b_\mathrm{u}(u)\leq \left(V(x_0)-x_0^\top P_\mathrm{uc}^* x_0\right)/\xi\right\}\right..  
\]
Here, $P_{\text{uc}}^*$ is the solution to the discrete-time algebraic Riccati equation related to the infinite-horizon LQR problem. Next, Eq.~\eqref{eq::suboptimality} in Appendix~\ref{app::m_bound} gives the suboptimality bound
$\|x_0^+-x_1^{\star}\|\leq c_{\mathrm{x}} \kappa^{\frac{\ell}{2}}\|x_0\|$ and similarly, we can have $\|u_0^\ell-u_0^{\star}\|\leq c_{\mathrm{u}} \kappa^{\frac{\ell}{2}}\|x_0\|$
with constants $c_{\mathrm{x}},c_{\mathrm{u}}>0$ such that combining the inequalities above yields the constrained violation bounds
\begin{subequations}
\begin{align}
&G^\mathrm{x}x_1^{\star} \leq d^\mathrm{x} + \hat z_\mathrm{x}(x_0,\delta)\\
\Rightarrow\;\; & \|G^\mathrm{x}x_0^+ -d^\mathrm{x}\|\leq \|\hat z_\mathrm{x}(x_0,\delta)\|+ \|G^\mathrm{x}(x_1^{\star} - x_0^+)\| \\
\Rightarrow\;\; & \|G^\mathrm{x}x_0^+ -d^\mathrm{x}\|\leq \|\hat z_\mathrm{x}(x_0,\delta)\|+ \bar c_{\mathrm{x}} \kappa^{\frac{\ell}{2}}\|x_0\|
\end{align}    
\end{subequations}
and an analogous derivation for the control input yields
\begin{align}
\|G^\mathrm{x}u_0^\ell -d^\mathrm{x}\|\leq \|\hat z_\mathrm{u}(x_0,\delta)\|+ \bar c_{\mathrm{u}} \kappa^{\frac{\ell}{2}}\|x_0\|
\end{align}
with $\bar c_{\mathrm{x}},\bar c_{\mathrm{u}}>0$.
Compared to the existing result based on the optimal solution, we have an additional terms  $\bar c_{\mathrm{x}} \kappa^{\frac{\ell}{2}}\|x_0\|$ and $\bar c_{\mathrm{u}} \kappa^{\frac{\ell}{2}}\|x_0\|$. It is clear that if we run an infinite number of iterations, i.e., $\ell\to\infty$, then the additional terms vanish.

Regarding the region in which there is no constraint violation, unfortunately, we cannot establish an analogous result from~\cite[Lemma~5]{RRLB-linear-MPC}. This is a direct consequence of the additional terms in the violation bound resulting from the suboptimality introduced by the finite number of iterations. Here, we only have the constraint violation in the proposed control scheme converging to zero for a fixed $\bar \ell$ following the asymptotical stability guarantee.

\subsubsection{Implementation Details}
After initialization and scaling, Algorithm~\ref{alg::aladin2} solves the unconstrained, convex NLPs from Step~2a) in parallel using a Newton method. Each local iterate is updated as $ y_{i,k}^{m+1} = y_{i,k}^{m} + \Delta y_{i,k}^m $ with $\Delta y_{i,k}^m = -\left( \nabla^2_{y_{i,k}}J_{i,k}(y_{i,k}^m) \right)^{-1} \nabla_{y_{i,k}}J_{i,k}(y_{i,k}^m)$, 
where $ J_{i,k}(y_{i,k}) $ denotes the twice continuously differentiable, convex decoupled problems~\eqref{eq::1stdeNLP}--\eqref{eq::NthdeNLP}, defined for $ i \in \mathcal{N} $ as
\begin{align*}
J_{i,0} &= \mathcal{L}_{i,0}(y_{i,0},\lambda_{\mathcal{N}_i,0}^\ell) + \Psi_{i,0}(y_{i,0} - \bar{y}_{i,0}^\ell),\\
J_{i,k} &= \mathcal{L}_{i,k}(y_{i,k},\lambda_{\mathcal{N}_i,k}^\ell, \lambda_{i,k-1}^\ell) + \Psi_{i,k}(y_{i,k} - \bar{y}_{i,k}^\ell),\\
J_{i,N} &= \mathcal{L}_{i,N}(y_{i,N}, \lambda_{i,N-1}^\ell) + \Psi_{i,N}(y_{i,N} - \bar{y}_{i,N}^\ell).
\end{align*}
The Newton method is run for a fixed number of iterations $ m_\text{T} $, as suggested in~\cite{Feller2018,BoyVan:ConOpt:04}, yielding $ y_{i,k} = y_{i,k}^{m_{\text{T}}} $ as input for Step~2b). Note that spatial decomposition leads to lower-dimensional Hessians, making inversion efficient. The coupled QP~\eqref{eq::cLQR} in Step~2b), structurally similar to an LQR problem, can be solved centrally via Riccati recursion~\cite{Bertsekas2012} if a global coordinator exists. 
Specifically, we compute: $\mathcal{P}_{k-1} = Q + A^\top\mathcal{P}_kA - (B^\top\mathcal{P}_kA)^\top R_\text{e}(B^\top\mathcal{P}_kA)$ with $R_{\textrm{e}} = (B^\top\mathcal{P}_kB + R)^{-1} $ and $ \mathcal{P}_N = \mathrm{diag}_{i\in\mathcal N} (\mathcal{P}_i ) $. The gains $ K_k = -R_{\textrm{e}}\mathcal{P}_kA $ yield the coupled updates $ \bar{y}_{i,k} $ and $ \Delta \lambda_{i,k} $. If only local communication is allowed, decentralized methods such as conjugate gradient methods can be used~\cite{Engelmann2020}. 
In either case, no external solver is required for Algorithm~\ref{alg::aladin2}.

\section{Illustrative Example}
\label{sec::case}
We demonstrate the proposed MPC scheme using a spring-damper-cart system with coupled inverted pendulums. The system connectivity graph is defined by \( \mathcal{N}_i = \{ i{-}1, i{+}1 \} \) for \( i = 2,\ldots,M{-}1 \), and \( \mathcal{N}_1 = \{2\} \), \( \mathcal{N}_M = \{M{-}1\} \). The state \( x_{i,k} = [p_{i,k}, \dot{p}_{i,k}, \varphi_{i,k}, \dot{\varphi}_{i,k}]^\top \) includes the position and velocities of the cart and pendulum, and the control input is the force \( u_{i,k} = F_{i,k} \). The system is linearized for controller design and simulation; full benchmark details are in~\cite{stomberg2025}.

Simulations were run in \texttt{MATLAB R2024a} on an 8-core AMD Ryzen 7 PRO with 16\,GB RAM. We compare centralized MPC with the proposed distributed MPC\footnote{
MPC setup: $Q_i = \text{diag}([1, 0.1, 10, 0.1])$, $R_i = 10^{-4}$, $P_i$ by~\eqref{eq::LMI}, sampling time $0.04$\,s, fixed $\delta = 10^{-4}$ of RRLB. The input and cart's speed constraints: $| \dot{p}_{i,k} | \leq 5$, $ | u_{i,k} | \leq 100$. Initialization: $x_{i,0} = [i, 0, \pi, 0]^\top$.}
(Algorithm~\ref{alg::aladin2}), using Newton's method capped at 2 iterations for solving decoupled NLPs numerically. The closed-loop performance achieved by Algorithm~\ref{alg::aladin2} yields error $|\sum_{t=0}^\infty l(x(t),u(t))-l(x^*(t),u^*(t))|/|\sum_{t=0}^\infty l(x^*(t),u^*(t))|$ less than 0.05\,\% with \( \bar{\ell} = 2 \). Furthermore, the constraints were active but never violated: inputs during the first two control steps and states from control steps 6 to 23.

\begin{center}
\scriptsize
\renewcommand{\arraystretch}{1.3}
\begin{tabular}{|c|c|cc|cc|cc|cc|}
\hline
\multirow{3}{*}{$M$} & \multirow{3}{*}{$N$} & \multicolumn{2}{c|}{\multirow{2}{*}{\texttt{GUROBI}}} & \multicolumn{2}{c|}{\multirow{2}{*}{\texttt{OSQP}}} & \multicolumn{4}{c|}{Algorithm~2} \\ \cline{7-10}
& & & & & & \multicolumn{2}{c|}{$\bar{\ell} = 2$} & \multicolumn{2}{c|}{$\bar{\ell} = 5$} \\ \cline{3-10}
& & $t_{\text{avg}}$ & $t_{\text{max}}$ & $t_{\text{avg}}$ & $t_{\text{max}}$ & $t_{\text{avg}}$ & $t_{\text{max}}$ & $t_{\text{avg}}$ & $t_{\text{max}}$  \\ \hline
5 & 40 & 20.2 & 26.6 & 7.4 & 61.2 & 8.3 & 10.7 & 18.8 & 25.5 \\
5 & 80 & 78.3 & 91.6 & 15.6 & 117.9 & 15.0 & 16.9 & 25.9 & 38.5 \\
8 & 40 & 80.9 & 105.9 & 27.5 & 139.9 & 10.3 & 13.8 & 27.3 & 36.2   \\
8 & 80 & 122.4 & 148.2 & 85.7 & 484.5 & 19.2 & 23.0 & 38.2 & 58.8 \\
\hline
\end{tabular}    
\end{center}

The table above shows runtimes (in [ms]) for multiple values of \( M \) and \( N \) for different solvers. Algorithm~\ref{alg::aladin2} yields competitive results, and one can see that introducing the tailored parallel computing framework accelerates the online computation effectively with problem scale-up. While \texttt{OSQP} is also fast, particularly in small-scale cases, it can allow violations of constraints, resulting in open-loop infeasibility. \texttt{Gurobi} and general-purpose Newton solvers were substantially slower due to the size and complexity of the problem. Note that the goal of the comparison was to demonstrate that the proposed method produces reliable closed-loop performance, achieves competitive runtimes, while remaining applicable in distributed settings, particularly under fixed-time computation constraints.

\section{Conclusion}
\label{sec::conclusions}

This paper presents a real-time MPC solver based on a relaxed recentered logarithmic barrier method for distributed optimization. Although the resulting control actions are suboptimal, closed-loop stability is guaranteed. The use of a fixed number of iterations per sampling time eliminates the need for stopping-criterion checks and reduces communication overhead compared to conventional distributed MPC schemes. Moreover, the solver’s coordination layer enables a straightforward, fully decentralized implementation following the approach in~\cite{Engelmann2020}.

\bibliographystyle{ieeetr}
\bibliography{reference}

\vspace{-0.2cm}
\appendices
\section{Proof of Theorem~\ref{Thm::globallinear}}
\label{app::convergence}
Based on Assumption~\ref{ass::aladin}\,a), we have there exists a positive constant $c_{\mathrm{L}}$ such that
\begin{equation}
\label{eq::Lipschitz}
\left\|\nabla f(y_1)-\nabla f(y_2)\right\|_2\leq c_{\mathrm{L}} \cdot \left\|y_1-y_2\right\|_{H}
\end{equation}
holds. Before we prove Theorem~\ref{Thm::globallinear}, we recall the result below.
\vspace{-0.8cm}
\begin{Lemma}[~{\cite[Theorem~1]{houska2017convex}}]
\label{eq::dscent_convergence}
Let Assumption~\ref{ass::aladin} hold,
the iterates of Algorithm~\ref{alg::ALADIN} satisfy the inequality
\begin{eqnarray}
\label{eq::Lyapunov}
\Omega( \bar y^{\ell+1}, \lambda^{\ell+1} ) \leq \Omega( \bar y^\ell, \lambda^\ell ) - 4 \Phi( y^\ell - y^\star) .
\end{eqnarray}
\end{Lemma}
\vspace{-0.12cm}
This lemma constructs the strict descent of $\Omega$ for the iterate of Algorithm~\ref{alg::ALADIN}. The interested reader is referred to~\cite{houska2017convex} for a more detailed proof of this lemma. Next, in order to establish the linear convergence condition~\eqref{eq::linearcontraction}, we show there exists a constant $\gamma$ such that 
\begin{equation}
\label{eq::inequality1}
\Omega( \bar y^{\ell+1}, \lambda^{\ell+1} )\leq \gamma^2  \cdot \Phi( y^\ell - y^\star).
\end{equation}
First, we recall the optimality conditions of~\eqref{eq::coQP} given by
\begin{subequations}
\label{eq::KKTcQP1}
\begin{align}
\lambda^{\ell+1} =\;& \lambda^\ell + \Sigma^{-1}(2D y^\ell - D \bar y^\ell -e),\\
\text{and}\quad\bar y^{\ell+1} =\;&2y^{\ell}-\bar y^\ell -H^{-1}D^\top(\lambda^{\ell+1}-\lambda^\ell)
\end{align}
\end{subequations}
with $\Sigma=DH^{-1}D^\top$.
Next, substituting the first-order optimality condition of~\eqref{eq::deNLP:general}, i.e.,
$\nabla f(y^\ell) + D^\top \lambda^\ell + H(y^\ell -\bar y^\ell)=0$
into~\eqref{eq::KKTcQP1} yields
\begin{subequations}
\label{eq::KKTcQP2}
\begin{align}\label{eq::KKTdual}
\lambda^{\ell+1} =\;& \Sigma^{-1}D(y^\ell - H^{-1}\nabla f(y^\ell)-e),\\\label{eq::KKTprimal} 
\text{and}\quad\bar y^{\ell+1} =\;&y^{\ell}-H^{-1}(\nabla f(y^\ell)+D^\top\lambda^{\ell+1}).
\end{align}
\end{subequations}
Now, we can bound the left-hand side of~\eqref{eq::inequality1} by $\|\lambda^{\ell+1} -\lambda^\star\|$
\begin{align*}
\overset{\eqref{eq::KKTdual}}{=}&
\left\|\Sigma^{-1}D(y^\ell -y^\star- H^{-1}(\nabla f(y^\ell)-\nabla f(y^\star)))\right\|_\Sigma\\[-0.1cm]
\leq&\left\|\Sigma^{-1}D(y^\ell -y^\star)\right\|_\Sigma+\left\| \Sigma^{-1}DH^{-1}(\nabla f(y^\ell)-\nabla f(y^\star))\right\|_\Sigma\\[-0.1cm]
\overset{\eqref{eq::Lipschitz}}{\leq}&\left\|H^{-\frac12}D^\top \Sigma^{-1} D H^{-\frac12}\right\|_2\left\|y^\ell-y^\star\right\|_{H}\\[-0.1cm]
+ &c_L\left\|H^{-1}D^\top\Sigma^{-1}DH^{-1}\right\|_2\left\|y^\ell-y^\star\right\|_{H}=\gamma_1\left\|y^\ell-y^\star\right\|_{H}    
\end{align*}
and $\left\|\bar y^{\ell+1}-y^\star\right\|_{H}$
\begin{align*}
\overset{\eqref{eq::KKTprimal}}{:=}&
\left\|y^{\ell}- y^\star-H^{-1}(\nabla f(y^\ell)-\nabla f(y^\star)+D^\top(\lambda^{\ell+1}-\lambda^\star))\right\|_{H}\\[-0.1cm]
\leq&\left\|y^\ell-y^\star\right\|_{H} + \left\|H^{-\frac12}\right\|_2\cdot \left\|(\nabla f(y^\ell)-\nabla f(y^\star))\right\|_2 \\[-0.1cm]
&+ \left\|\Sigma^{-\frac12}DH^{-1}D^\top\Sigma^{-\frac12}\right\|_2\left\|\lambda^{\ell+1}-\lambda^2\right\|_\Sigma\\[-0.1cm]
\leq& \gamma_2\cdot \left\|y^\ell-y^\star\right\|_{H}\quad \text{with}\\[-0.1cm]
\gamma_1&= \left\|H^{-\frac12}D^\top \Sigma^{-1} D H^{-\frac12}\right\|_2+ c_L\left\|H^{-1}D^\top\Sigma^{-1}DH^{-1}\right\|_2,\\[-0.1cm]
\gamma_2&=1+ \left\|\Sigma^{-\frac12}DH^{-1}D^\top\Sigma^{-\frac12}\right\|_2\hspace{-2mm}+ \gamma_1\left\|\Sigma^{-\frac12}DH^{-1}D^\top\Sigma^{-\frac12}\right\|_2.
\end{align*}
Then, we sum both estimates such that 
$\Omega(\bar y^{\ell+1},\lambda^{\ell+1}) = \left\|\bar y^{\ell+1}-y^\star\right\|_{H}^2 + \left\|\lambda^{\ell+1}-\lambda^\star\right\|_\Sigma^2
\leq \gamma^2\left\|y^\ell-y^\star\right\|_{H}^2 = \gamma^2 \Phi(y^\ell -y^*)
$ with $\gamma^2=\gamma_1^2+\gamma_2^2$.
Finally, substituting it into~\eqref{eq::Lyapunov} yields
$
\Omega(\bar y^{\ell+1} ,\lambda^{\ell+1})\leq \kappa\cdot \, \Omega(\bar y^\ell, \lambda^\ell)$ with $\kappa=\gamma^2/(4+\gamma^2)<1$,
which concludes the proof. \phantom{.} \hfill$\square$

\section{Proof of Theorem~\ref{thm::nominalstability}}
\label{app::nominal}
As discussed in Remark~\ref{rem::nominal}, the main idea of constructing the closed-loop system stability of~\eqref{eq::rrbMPC} is to guarantee the terminal cost $L^{\mathrm{f}}(\cdot)$ is a Lyapunov function. Following Proposition~\ref{prop::Lyapunov_terminal}, the method proposed to the separable terminal cost design guarantees $L^{\mathrm{f}}(\cdot)=\sum_{i=1}^ML^{\mathrm{f}}_i(\cdot)$ is a Lyapunov function for~\eqref{eq::dynamic}, i.e., conditions~\eqref{eq::Lyapunov_term} hold. This implies we have 
\begin{align*}
&V(x_{1}^\star)-V(x_0)+\sum_{i=1}^M L_i(x_{i,0},u_{i,0}^\star)\\[-0.16cm]
\leq\;\, &\sum_{i=1}^M\left(L^{\mathrm{f}}_i(x_{i,N}^+)- L^{\mathrm{f}}_i(x_{i,N}^\star) + L_i(x_{i,N}^\star,K_{\mathcal N_i}x_{\mathcal N_i,N}^\star) \right) \\[-0.1cm]
\overset{\eqref{eq::extendedstage}}{\leq}\; & \sum_{i=1}^M\left(L^{\mathrm{f}}_i(x_{i,N}^+)- L^{\mathrm{f}}_i(x_{i,N}^\star) + \widetilde l_i(x_{i,N}^\star,K_{\mathcal N_i}x_{\mathcal N_i,N}^\star) \right)\\[-0.1cm]
\overset{\eqref{eq::localLyapunov}}{\leq} & \sum_{i=1}^M \vartheta_i(x_{\mathcal N_i,N}^\star)\overset{\eqref{eq::residue}}{\leq} 0
\end{align*}
with \( x_{1}^\star = A x_{0} + B u_{0}^\star \), representing state evolution under the optimal control action, and \( x_{i,N}^+ = (A_{\mathcal{N}_i} + B K_{\mathcal{N}_i}) x_{\mathcal{N}_i,N}^\star \), denoting state evolution under the stabilizing feedback control law. The inequality above indicates that $V(\cdot)$ is a Lyapunov function, which is sufficient to prove asymptotic stability. \hfill$\square$

\section{Proof of Lemma~\ref{Lem::value}}
\label{app::Lem1}
First, by eliminating the dynamic equations using the single shooting method, we write down the value function $V(\cdot)$ given by~\eqref{eq::mpc_dense} as an unconstrained parametric optimization problem $V(x_0):=\min_u \;\mathcal F(u,x_0)$
and the solution map $u^\star(\cdot)$ is determined by the initial state $x_{0}$. Based on its first-order optimality condition, we have $\nabla_u \mathcal F(u^\star(x_0),x_0)=0
\Rightarrow$
\[
\nabla_{uu} \mathcal F(u^\star(x_0),x_0) \frac{ \mathrm{d} u^\star(x_0)}{ \mathrm{d} x_0} + \nabla_{ux_0} \mathcal F(u^\star(x_0),x_0)=0  
\]
hold for any $x_0\in\mathbb R^{n_\mathrm{x}}$. Then, as the function $\mathcal F(\cdot,\cdot)$ is twice continuously differentiable for both arguments, and its Hessian matrix is lower and upper bounded, such that we have the solution $u^\star(\cdot)$ is first-order Lipschitz continuously differentiable, i.e., there exists a constant $c_u>0$ such that 
\begin{equation}
\label{eq::Lipschitz_y}
\|u^\star(a)-u^\star(b)\|\leq c_u\cdot \|a-b\|
\end{equation}
holds for any $a,b\in\mathbb R^{n_\mathrm{x}}$, which follows Implicit Function Theorem~\cite{bonnans2013perturbation}. Then, based on the Lipschitz continuity of $u^\star(\cdot)$, we can further have $|V(x_0^+)- V(x_1^\star)|\leq$
\begin{align}\label{eq::value1}
\tilde\eta_1\|\nabla V(x_1^\star)\|\|x_0^+-x_1^\star\|_Q + \eta_2 \|x_0^+-x_1^\star\|_Q^2 
\end{align}
As $\nabla V(0)=0$ holds and $\nabla V(\cdot)=\nabla_{x_0} \mathcal F(u,\cdot)$ is Lipschitz continuous, we can have 
\begin{equation}
\label{eq::value2}    
\|\nabla V(x_1^\star)\|\leq c_v\|x_1^\star\|\leq c_L\|x_0\|_Q\;\;\text{with}\;c_v,c_L>0,
\end{equation}
where the second inequality follows that $x_1^\star = A x_0 + Bu^\star(x_0)$ and $u^\star(\cdot)$ is Lipschitz continuous. Then, we can conclude the proof by combining~\eqref{eq::value1} and~\eqref{eq::value2} with $\eta_1=\tilde\eta_1\cdot c_L$.
\hfill$\square$

\section{Proof of Theorem~\ref{thm::m_bound}}
\label{app::m_bound}
The proof of Theorem~\ref{thm::m_bound} begins with the stationary condition of the coupled problem~\eqref{eq::cLQR} in form
\begin{equation}
\label{eq::SC_cQP}
H(\bar{y}^{\ell+1}-2y^{\ell}+\bar{y}^{\ell})+\mathcal{D}^\top\Delta\lambda^{\ell} = 0 \, ,
\end{equation}
with $\Delta\lambda^{\ell} = \lambda^{\ell+1}-\lambda^{\ell}$ and $\mathcal{D} = $
\begin{equation*}
\nabla_{\Delta\lambda,\bar{y}} \left( \sum_{k=0}^{N-1}\Delta\lambda_{i,k}^\top\sum_{j\in\mathcal{N}_i}\mathcal{A}_{i,j,k}\bar{y}_{i,k} + \sum_{k=1}^{N}\Delta\lambda_{i,k-1}^\top\mathcal{C}_{i,k}\bar{y}_{i,k}\right) \, .
\end{equation*}
Since $H\succ 0$, we can derive decoupled solution $y^\ell$ as
\begin{equation}
\label{eq::dQP_sol}
y^\ell = \frac{1}{2} H^{-1}\mathcal{D}^\top(\lambda^{\ell+1}-\lambda^{\ell})+\frac{1}{2}(\bar{y}^{\ell+1}+\bar{y}^{\ell}) \, .
\end{equation}
Then, we have $x_0^+-x_1^\star = B(y_0^{\ell}-y_0^\star) = \mathcal{S}(y^{\ell}-y^\star)$ with
$\mathcal{S} = [B,0,\ldots,0]$, that is further derived using~\eqref{eq::dQP_sol} as
\begin{align*}
x_0^+-x_1^\star& = \mathcal{S}\left(\frac{H^{-1}\mathcal{D}^\top(
\lambda^{\ell+1}-\lambda^{\ell})}{2} + \frac{\bar{y}^{\ell+1}+\bar{y}^{\ell}}{2}-y^\star\right) \\[-0.1cm]
&=\frac{1}{2}\mathcal{S}H^{-1}\mathcal{D}^\top(\lambda^{\ell+1}-\lambda^{\star}) + \frac{1}{2}\mathcal{S}(\bar{y}^{\ell+1}-y^\star) \\[-0.1cm]
&\quad+ \frac{1}{2}\mathcal{S}H^{-1}\mathcal{D}^\top(\lambda^{\star}-\lambda^{\ell}) + \frac{1}{2}\mathcal{S}(\bar{y}^{\ell}-y^\star) \, .
\end{align*}
Moreover, there exits a constant \(\sigma > 0\) such that \(B_i^\top Q_i B_i \preceq \sigma R_i\) holds for each \(i \in \mathcal{N}\). This directly implies that \(\sigma\) also satisfies the inequality \(\mathcal{S}^\top Q\mathcal{S} \preceq \sigma H\).
As a result, we have 
\begin{align}\notag
&\|x_0^+-x_1^\star\|_Q^2\leq (\lambda^{\ell}-\lambda^{\star})^\top \mathcal{D} H^{-1}\mathcal{S}^\top Q \mathcal{S}H^{-1}\mathcal{D}^\top (\lambda^{\ell}-\lambda^{\star})\\[-0.1cm]\notag
&+(\lambda^{\ell+1}-\lambda^{\star})^\top \mathcal{D} H^{-1}\mathcal{S}^\top Q \mathcal{S}H^{-1}\mathcal{D}^\top (\lambda^{\ell+1}-\lambda^{\star}) \\[-0.1cm]\label{eq::suboptimality}
&+ (\bar{y}^{\ell+1}-y^\star)^\top \mathcal{S}^\top Q \mathcal{S}(\bar{y}^{\ell+1}-y^\star)  \\[-0.1cm]\notag
&+  (\bar{y}^{\ell}-y^\star)^\top \mathcal{S}^\top Q \mathcal{S}(\bar{y}^{\ell}-y^\star) \\[-0.1cm]\notag
&\leq \sigma\Omega(\bar{y}^{\ell+1},\lambda^{\ell+1}) + \sigma\Omega(\bar{y}^{\ell},\lambda^{\ell})
\leq 2\sigma(1+\kappa)\kappa^{\ell-1}\Omega(\bar{y}^1,\lambda^1) . 
\end{align}
Based on Assumption~4 and the rescaling Step 1) in Algorithm~\ref{alg::aladin2}, we have $\Omega(\bar{y}^1,\lambda^1) \leq 2\alpha^2\|x_0\|^2_Q$ such that 
$$\|x_0^+-x_1^\star\|_Q^2 \leq 4\sigma(1+\kappa)\kappa^{\ell-1}\alpha^2\|x_0\|^2_Q.$$ 
Then, based on Lemma~\ref{Lem::value}, we have $$|V(x_0^+)-V(x_1^\star)|\leq\eta_1 \|x_0\|_Q\|x_0^+-x_1^\star\|_Q+\eta_2 \|x_0^+-x_1^\star\|_Q^2.$$
Combining two inequalities above yields $|V(x_0^+) -V(x_1^\star)|\leq\left(2\eta_1\alpha\sqrt{\sigma(1+\kappa)/\kappa}
+4\eta_2\alpha^2\sigma(1+\kappa)/\kappa\right)\kappa^{\nicefrac{\ell}{2}}\|x_0\|^2_Q$. Here, recalling that $\kappa < 1$, we have $\kappa^{\nicefrac{\bar \ell}{2}} > \kappa^{\bar{\ell}}$.
Then, it is sufficient to satisfy the Lyapunov descent condition from Theorem~\ref{thm::nominalstability} in the form
$V(x_0^+) - V(x_0)  \leq  V(x_0^+) - V(x_1^\star) - \|x_0\|^2_Q 
\leq(\beta_1-1)\|x_0\|^2_Q < 0$,
which is achieved by enforcing $\beta_1<1$:
\begin{align}\notag
&\beta_1 = \left(2\eta_1\alpha\sqrt{\sigma(1+\kappa)/\kappa}
+ 4\eta_2\alpha^2\sigma(1+\kappa)/\kappa\right)\kappa^{\nicefrac{\ell}{2}} < 1\\[-0.1cm]\label{eq::l_bar_x0_u}
\hspace{-1mm}\Rightarrow  &\bar{\ell} > \frac{2\log\left(2\eta_1\alpha\sqrt{\sigma(1+\kappa)/\kappa}+4\eta_2\alpha^2\sigma(1+\kappa)/\kappa\right)}{\log(1/\kappa)}
\end{align}
to guarantee asymptotic stability. \hfill$\square$

\end{document}